\documentclass[aip,jmp, amsmath,amssymb,reprint,]{revtex4-1}

\usepackage{graphicx}
\usepackage{dcolumn}
\usepackage{bm}
\usepackage{color}

\begin{document}
\title{Continuous-frequency measurements of high-intensity microwave electric fields with atomic vapor cells}

\author{D.~A.~Anderson}
\email{dave@rydbergtechnologies.com.}
\affiliation{Rydberg Technologies LLC, Ann Arbor, Michigan 48104 USA}
\author{G.~Raithel}
\affiliation{Rydberg Technologies LLC, Ann Arbor, Michigan 48104 USA}
\affiliation{Department of Physics, University of Michigan, Ann Arbor, Michigan 48109 USA}

\date{\today}

\begin{abstract}
We demonstrate continuous-frequency electric field measurements of high-intensity microwaves via optical spectroscopy in a small atomic vapor cell.  The spectroscopic response of a room-temperature rubidium atomic vapor in a glass cell is investigated and employed for absolute measurements of K$_a$-band microwave electric fields from $\sim$200~V/m to $>$1~kV/m over a continuous frequency range of $\pm $1~GHz (15\% band coverage).  It is established that in strong microwave fields frequency-specific spectral features allow for electric field measurements over a large continuous frequency range.
\end{abstract}

\maketitle
\section{Introduction}
Significant progress has been made in atom-based measurements of microwave electric fields using electromagnetically induced transparency in room-temperature alkali metal vapors, in which a laser beam addresses highly field-sensitive atomic Rydberg states~\cite{Mohapatra.2007, Sedlacek.2012, Holloway.2014}.  The Rydberg-atom-based measurement approach has garnered broad interest at national metrology institutes for the establishment of new atomic measurement standards of microwave and millimeter-wave electric fields~\cite{Gordon.2014}, and holds promise for the development of atomic microwave electric-field sensors and measurement technologies with unprecedented bandwidth and dynamic range.

Measurements of high-intensity electric fields present a considerable challenge to existing sensor technologies.  In high-power measurement applications, a main limitation of existing antenna probes is their susceptibility to damage from field-induced heating of conductive elements, which can occur at even relatively modest kV/m fields.  The heating can also lead to un-repeatable drift behavior due to field-induced mechanical and electrical changes in the probing element.  Atom-based sensors, which contain no metal or conductive materials, are effectively immune to electromagnetic heating effects, with damage thresholds dictated instead by either ionization of the atoms or the dielectric strength of the glass vapor cell (which can be in the MV/m range). Atom-based probes thereby afford precise and repeatable absolute-standard field measurements over a wide measurement range, without drift or damage.  In recent work, the Rydberg-atom-based measurement method was extended to fields up to $\sim$300~V/m, demonstrating electric ($E$) field measurements of resonant high-frequency $K_u$-band microwaves~\cite{Anderson.2016} as well as off-resonant long-wavelength (10 to 500~MHz) radio-frequency fields~\cite{Bason.2010, Miller.2016,Jiao.2016,Veit.2016}.  Towards applications, it is desirable to extend the method's capability to measurements of microwave fields above 1~kV/m.  The accuracy has been limited in part by microwave field inhomogeneities inside the spectroscopic cells.  In turn, this has precluded the ability to test atomic-physics models for strong atom-field interactions that are required to quantify strong microwave electric fields.

Another outstanding challenge with the atom-based approach has been to achieve continuous microwave frequency measurement capability.  To date, the technique has relied on dipole-allowed resonant or near-resonant transitions between Rydberg states.  Due to their large transition dipole moments ($\sim 10^3 - 10^4 ~ea_0$), these transitions afford a maximal atomic response and sensitivity to microwave electric fields.  In exchange for the high sensitivity, however, the reliance on near-resonant transitions restricts any measurement of microwave fields to a discrete set of atomic transition frequencies.  While many Rydberg transitions are accessible, spanning a large (GHz to THz) frequency range~\cite{Holloway2.2014,Miller.2016,Wadearxiv.2016}, true continuous-frequency measurement capability requires measurements of microwave fields that are off-resonant with any atomic transition.  This presents a significant practical limitation of the atom-based field sensing approach as demonstrated to-date. While microwave field measurements for small ($\sim$10~MHz) frequency detunings around atomic transitions have been demonstrated using off-resonant Autler-Townes splittings~\cite{Simons.2016}, measurements for larger continuous frequency detunings have yet to be demonstrated.

In this work we demonstrate high-intensity microwave field measurements exceeding 1~kV/m and strong-field measurements over a continuous microwave frequency range in the $K_a$-band, up to $\pm 1$~GHz detuned from the next relevant atomic transition.  In section I, we describe the experimental setup.  In section II, spectral maps for the $45D-46D$ two-microwave-photon transition in rubidium are presented and described. This component of the work is focused on validation of the utilized atom-field coupling model in strong, near-resonant microwave fields, ranging from $\sim$200~V/m to 1~kV/m.  In section III, we focus on extending atom-based microwave field measurements into the realm of far detuning. In the presented data, we employ the same transition as in Section~II, but the microwave frequency is substantially detuned.  In section IV we describe on-going and future work.

\section{Experimental setup}
\label{sec:2}
\begin{figure}[h]
\includegraphics[width=8.6cm]{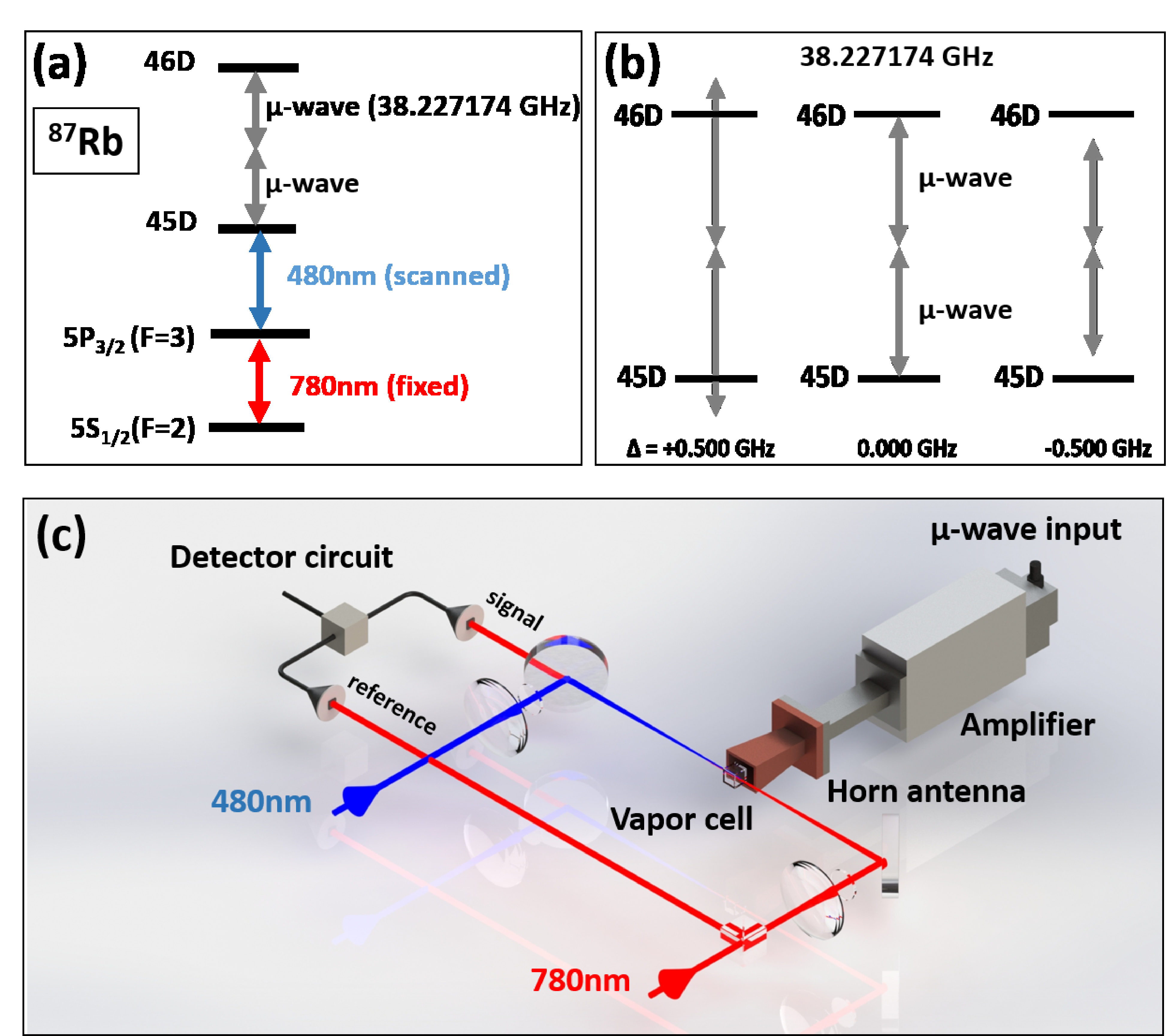}
\caption{(a) Energy-level diagram for the Rydberg EIT ladder scheme and field-free resonant two-photon microwave transition.  (b) Energy-level diagram showing microwave frequency tuning around field-free two-photon resonance. (c) High-intensity microwave electric field measurement platform.}
\label{fig:1}
\end{figure}

The experimental setup and the $^{87}$Rb Rydberg electromagnetically-induced transparency (EIT) ladder system used in the measurements are illustrated in Fig.~\ref{fig:1}.  The setup is shown in Fig.~\ref{fig:1}(c) and consists of an isotopic $^{87}$Rb cubic vapor cell with a 4~mm inner dimension and 1~mm-thick walls and two lasers at wavelengths $\lambda_p=$780~nm (probe) and $\lambda_c=$480~nm (coupler) counter-propagating through the center of the cell.  A reference beam is split off from the probe laser beam prior to entering the cell.  The EIT and reference 780~nm beams are detected on separate photodiodes and their signals subtracted to eliminate any probe laser power drifts.  The EIT probe beam is focused to a full-width-half-maximum (FWHM) of 70~$\mu$m through the cell and has a power of 5~$\mu$W.  The probe laser is frequency locked to the $\vert 5S_{1/2}, F=2\rangle\rightarrow\vert 5P_{3/2}, F=3\rangle$ transition.  The coupler laser is overlapped with the probe and focused to a FWHM of 85~$\mu$m in the cell and has a power of about 40~mW.  The coupler laser is scanned over a range of approximately 3~GHz around the $\vert 5P_{3/2}, F=3\rangle\rightarrow\vert Rydberg \rangle$ transition.  A small fraction of the coupler beam is sent through a Fabry-Perot cavity with a free spectral range of 1~GHz. The transmission of the coupler through the cavity is recorded simultaneously with the EIT spectra and used to correct for small frequency drifts of the coupling laser during the measurements.

A WR-28 horn antenna with 10~dB gain and interior dimensions of $8.00\times 10.67$~mm is positioned in front of the vapor cell.  Microwaves are generated by a signal generator and injected into a 40~dB amplifier that feeds into the horn.  The atomic response for a given injected microwave power is measured optically using EIT by scanning the coupler laser frequency and monitoring the absorption of the probe transmission through the cell.  Spectral maps of the atomic response are obtained at fixed microwave frequency as a function of input microwave power.  The maps are assembled as two-dimensional plots showing probe transmission versus microwave power and coupler-frequency detuning from the field-free EIT resonance.

\section{K$_a$-band microwave field measurements exceeding 1~kV/m}
\label{sec:3}
\begin{figure}[h]
\includegraphics[width=7cm]{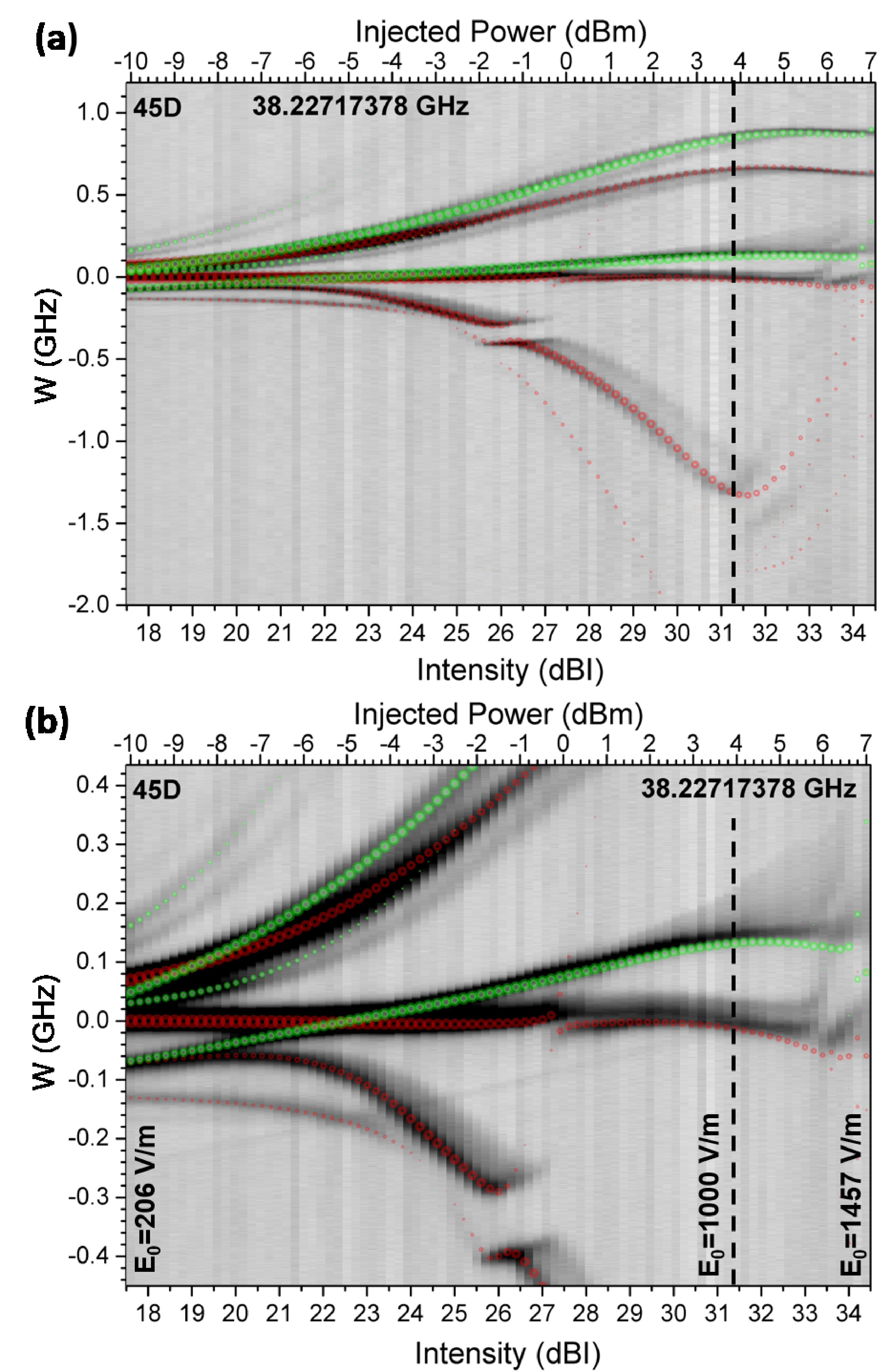}
\caption{(a) Experimental spectral map showing transmission of the probe laser (gray scale) versus coupler-laser frequency detuning, $W$, and injected microwave power (top axis).  The overlaid Floquet calculation shows excitation rates (circle areas) into the $m_j$=1/2 (red) and $m_j$=3/2 (green) Rydberg-state manifolds. The intensity $I$ in the calculation is given in dBI=10$\times \log(I$/W/m$^2)$ (bottom axis). (b) Crop of spectral map in (a) over a smaller coupler-laser scan range.}
\label{fig:2}
\end{figure}

Experimental spectral maps are shown in Fig.~\ref{fig:2} for 38.22717378~GHz microwaves driving the 45D$_{5/2}$-46D$_{5/2}$ two-photon transition for microwave powers injected into the amplifier ranging from -10 to 7~dBm in steps of 0.2~dBm.  The corresponding zero-field energy-level diagram is shown in Fig.~\ref{fig:1}(a). Compared to previous work~\cite{Anderson.2016}, the cell length scale ($d_{cell}$) is reduced by more than a factor of six and approaches the microwave half-wavelength ($\lambda_{MW}/2 \approx d_{cell}$).  This eliminates the possibility of any field inhomogeneities in the cell due to weak higher-order standing waves from the dielectric cell acting as a low-finesse etalon.  Further, since the optical beam sizes are small compared to the microwave half-wavelength, and their path through the center of the cell is perpendicular to the incident microwave field, field inhomogeneities in the spectra due to the fundamental harmonic standing wave are minimal.  These aspects afford higher spectral resolution and accuracy in the electric field determination. The experimental maps in Fig.~\ref{fig:2} are overlaid with a calculated Floquet map, from which the absolute intensity and the $E$-field are obtained.  The Floquet method is non-perturbative and is described in our previous work~\cite{Anderson.2016}.  The experimental and calculated spectral maps are in excellent agreement over nearly the entire range, exceeding field amplitudes $E_0=$1~kV/m (31.3~dBI).

The transition between weak-field and strong-field regimes occurs at lower field strengths than those considered in the present work and has been investigated in detail in previous work~\cite{Anderson.2014}.  In the weak-field regime, the microwave Rabi frequency is given by the frequency splitting between pairs of Autler-Townes peaks (typically below 100~MHz~\cite{ Sedlacek.2012, Holloway.2014}). Over the field range studied in the present work, single-photon Rabi frequencies between the relevant $D$- and nearby $P$ and $F$ states range from several GHz to several tens of GHz, approaching the microwave frequency. The two-photon Rabi frequencies range from several 100~MHz to several GHz. These numbers show that the present work is in a strong-field field regime, which requires Floquet theory in a large state space. The strong-field regime is qualitatively different from the low-field (Autler-Townes) regime; in the latter a two-level treatment is sufficient to describe most of the physics.  The observable for the field strength in the high-intensity spectra lies in the splitting of the zero-microwave-field EIT peak into a set of peaks, or Floquet states, with distinct energy-level shifts, separations, and line strengths.  This is a generalization of weak-field measurements based on pairs of Autler-Townes peaks; the generalization relies on exploiting the response of the atom over wide microwave field and frequency ranges desired in practical applications.  An implementation of a field measurement entails calculating a set of atomic spectra, and finding the best match between the measured spectrum obtained in an arbitrary field of interest with one of the calculated spectra.  The matching process, which could involve cross correlations between the spectra, yields field amplitudes and frequencies.  As seen in Fig.~\ref{fig:2}(a) and in the enlargement in Fig.~\ref{fig:2}(b), the Floquet states exhibit spectral features that include (1) a high density of states, (2) varying differential dynamic dipole moments throughout, and (3) multipole avoided crossings.  The combination of these attributes and $E$-field markers allows for strong-field measurement over the entire field range. Complementing the data with measurements near other Rydberg transitions would yield additional spectroscopic information that could be used to enhance the field measurement.

In Fig.~\ref{fig:2}(b) we note that at microwave fields above 1~kV/m the upper $m_j=3/2$ line of the central pair of spectral lines exhibits a continuous broadening to the blue by several hundred MHz, up to the highest microwave intensities.  This state-dependent feature may be attributable to state-dependent Rydberg-Rydberg interactions, electric fields of nearby ions, or a differential  susceptibility of the Floquet levels to microwave ionization.  From an applications standpoint, the deviations do not preclude measurement capability of fields exceeding 1~kV/m because Rydberg states with lower principal quantum numbers, which have a reduced response to microwave fields and are less prone to interactions, are readily available.

\section{Continuous-frequency microwave field measurements in the strong-field regime}

Weak microwave fields become detectable in Rydberg-atom-based EIT measurement via microwave-induced distortion and Autler-Townes splitting of EIT lines~\cite{Sedlacek.2012, Holloway.2014, Gordon.2014}. In this weak-field regime, microwave fields that are more than tens of MHz off-resonance from a Rydberg transition become hard detect, because AC level shifts decrease with increasing detuning. In contrast, in the strong-field regime, which is characterized by off-resonant microwave-induced mixing and higher-order shifts of Rydberg levels, even far-off-resonant microwave fields are detectable.  In the strong-field regime a multitude of Rydberg states becomes optically accessible, because of microwave-induced admixtures of $S$- or $D$-state character into many Rydberg states. Since the Rydberg transition driven by the EIT coupling laser beam (the 480-nm beam in Fig.~1~(a)) is subject to the usual electric-dipole selection rules, the microwave-admixed $S$- or $D$ character makes additional Rydberg states accessible to EIT measurements. This implies that all Rydberg levels with sufficient microwave-admixed $S$- or $D$-state character can be probed via Rydberg-EIT and can be exploited for atom-based microwave $E$-field measurements.

\begin{figure}[h]
\includegraphics[width=8cm]{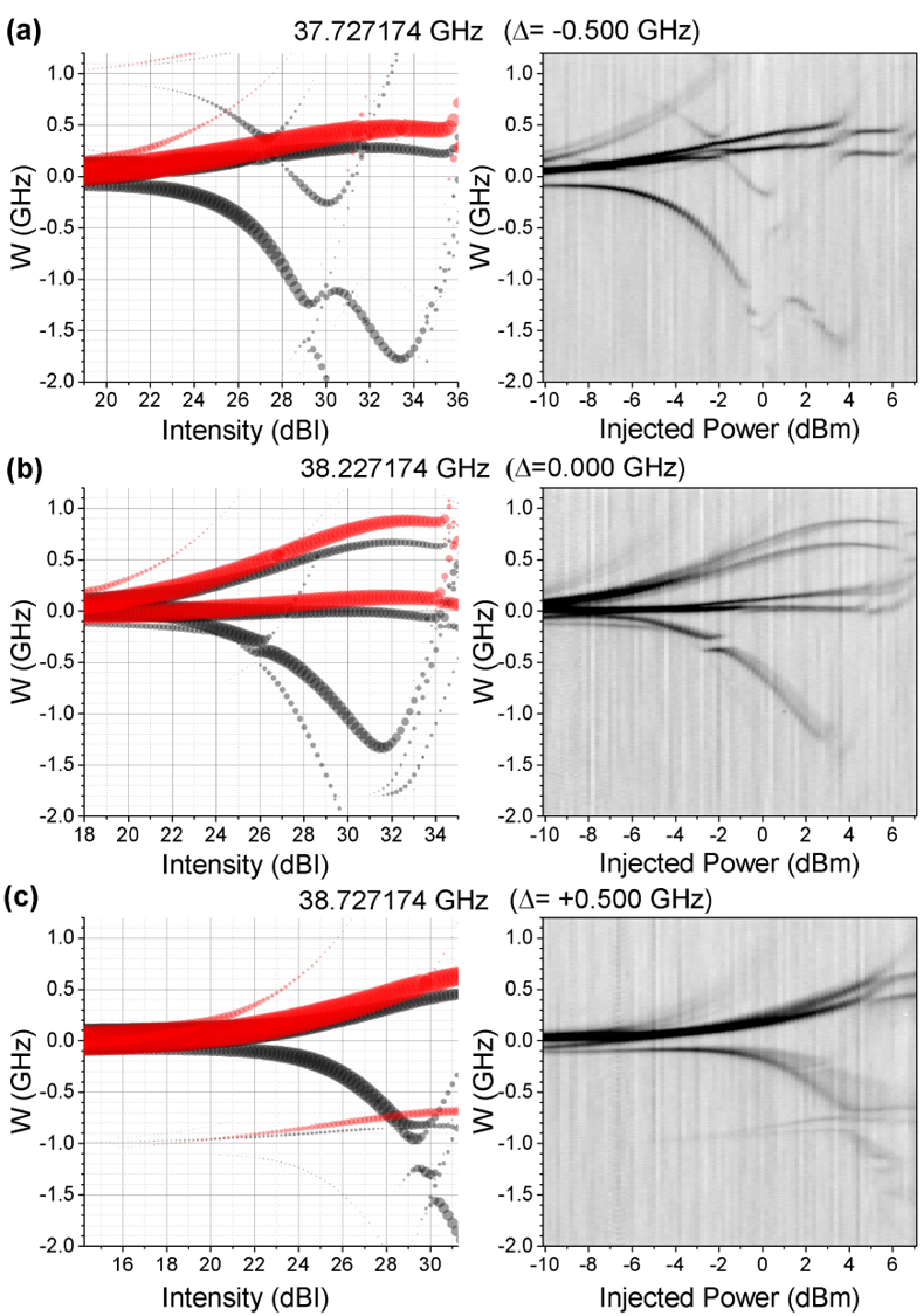}
\caption{Experimental and calculated spectral maps for (single-photon) microwave frequency detuning ($\Delta$) of (a) $-0.500$~GHz, (b) 0.000~GHz, and (c) +0.500~GHz from the 45D-46D two-photon transition. Experimental maps (right) are given as a function of injected microwave power in dBm.  The signal strength is represented on a linear gray scale in arbitrary units of the probe transmission. Calculated maps (left) with excitation rates to $m_j$=1/2 (black) and $m_j$=3/2 (red) states proportional to circle areas.}
\label{fig:3}
\end{figure}

\begin{figure*}[t]
\includegraphics[width=14cm]{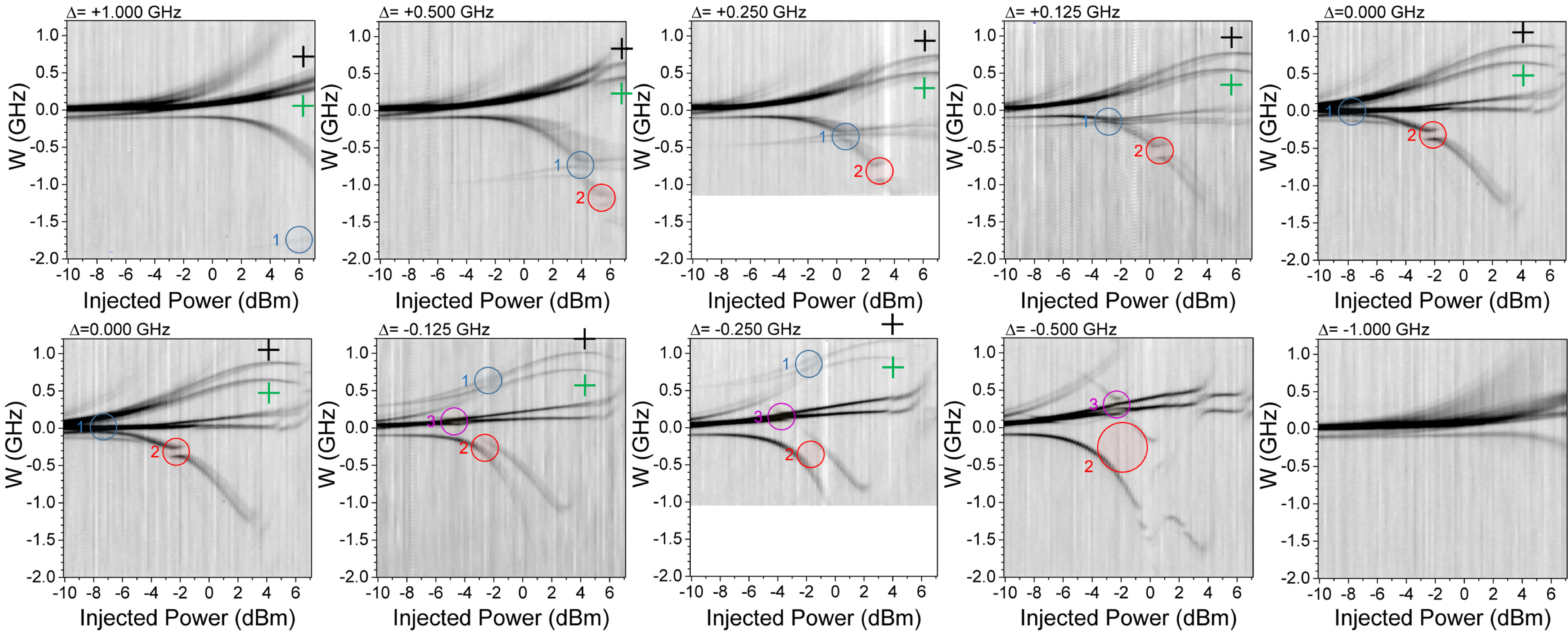}
\caption{Experimental spectral maps for detunings $\Delta$=+1.000, +0.500, +0.250, +0.125, and 0.000~GHz (top row) and $\Delta$=0.000, -0.125, -0.250, -0.500, and -1.000~GHz (bottom row), relative to the single-microwave-photon frequency for the two-photon $45$D to $46$D resonance condition at 38.227174~GHz.  The symbols on top of the data mark spectral features that aid in correlating microwave powers with electric fields over a wide and continuous frequency range.}
\label{fig:4}
\end{figure*}

To demonstrate continuous-frequency tuning with the Rydberg-atom-based approach in the strong-field regime, we obtain $K_a$-band microwave spectral maps for microwave frequencies ranging from 37.227174~GHz to 39.227174~GHz, $\pm$1~GHz around the 38.227174~GHz microwave frequency used for the resonant two-microwave-photon 45D to 46D transition investigated in section~\ref{sec:3}.  Figure~\ref{fig:1}(b) shows energy-level diagrams illustrating the microwave frequency change and atomic transition of interest. Figure~\ref{fig:3} shows the corresponding experimental and calculated spectral maps for (single-photon) microwave frequency detunings of $\Delta=$-0.500, 0.000, and +0.500~GHz from the two-photon 45D to 46D resonance. The experimental and calculated maps are in excellent agreement over nearly the entire power range.  Due to the substantial microwave-induced state mixing in the strong atom-field interaction regime, microwave-frequency-specific spectral features can be optically detected and used to determine the associated field amplitude for microwave fields far-off-resonance from any Rydberg transition.  We observe that a $\pm$0.500~GHz detuning of the microwave frequency from resonance has a pronounced effect on the features, resulting in spectral maps unique to each off-resonant microwave frequency, with distinct level crossings and shifts providing excellent $E$-field markers at these frequencies.  The spectral sensitivity to the exact microwave field value is maintained over the entire range of frequencies studied.

The response of the atomic system to microwave-frequency changes varies on a fine frequency scale, as one might expect from the dramatic changes between the panels shown in Fig.~\ref{fig:3}, and it extends to larger microwave detunings.  In Fig.~\ref{fig:4} we show experimental spectral maps for $\Delta=$+1.000, +0.500, +0.250, +0.125, 0.000, -0.125, -0.250, -0.500, and -1.000~GHz relative to the (single-photon) microwave frequency for the two-photon $45$D to $46$D resonance condition at 38.227174~GHz.  This corresponds to a $15\%$ coverage of the $K_{a}$ microwave band.  By comparing the series of spectral maps, one can identify how level shifts and avoided crossings move through the spectral map as a function of microwave frequency.  In the Fig.~\ref{fig:4} we overlay several types of symbols that identify Floquet-energy-level extrema and avoided crossings as they track through the maps with microwave frequency.  Crosses identify energy-level extrema and circles avoided crossings.  Circles with equal numbers mark avoided crossings that continuously transform into each other as the microwave frequency is changed.  These systematic shifts of lines and crossings provide excellent features for field determination. Following this approach, continuous multi-band frequency- and field-resolving measurements, spanning a frequency range from tens of GHz down to the MHz level away from any given transition, can be obtained. The requirement for the procedure to work is that the microwave field  must be strong enough to provide field-induced state mixing over the frequency range of interest.

Close inspection of Figs.~\ref{fig:3} and~\ref{fig:4} reveals, for $\Delta \ne 0$, that some of the faint Floquet lines are offset from the strongest lines by about $- 2 \Delta$.  These faint lines resemble “weak” Autler-Townes components of the 45D to 46D two-photon transitions. At $\Delta=0$, instead of faint, off-set Floquet lines one observes split, strong Floquet line pairs of approximately equal strength. Those resemble resonant low-field Autler-Townes pairs. Note, however, that the Floquet-level shifts, avoided crossings and non-linearities that are also measured are beyond the descriptive power of the low-field Autler-Townes model.

\section{Conclusion}
In this work we have demonstrated measurement capability using the Rydberg-atom-based field measurement approach to fields exceeding 1~kV/m.  A small spectroscopic cell was implemented to minimize field inhomogeneities in the measurement volume to obtain high spectral resolution and improved accuracy in field determination.  Spectroscopy in the strong-field regime was used to obtain field measurements for microwaves near- and far-off-resonant with the 45D-46D two-photon Rydberg transition, with detunings ranging up to $\pm 1$~GHz.  While we have selected a set of nine microwave frequencies to demonstrate measurements of off-resonant fields, the method is applicable to any value within the frequency continuum.

The ability to quantify the field using the Rydberg-atom-based field measurement approach relies on the existence of optically accessible, field- and frequency-sensitive spectral features. In the strong-field regime, this condition can be met over wide, continuous ranges of the microwave frequency.  In addition, the measurements can be conducted drawing from a selection of Rydberg levels with different principal quantum numbers, $n$, with different and overlapping useful frequency ranges.  This way it is possible to measure fields at arbitrary frequencies, within a range $\sim$1~MHz to $ > 500$~GHz.  Further, due to the $\sim n^2$ scaling of the Rydberg-Rydberg dipole matrix elements and the increasing density of Floquet states at higher microwave intensities, field amplitudes of tens of kV/m and above could be measured for an arbitrary microwave frequency by optical interrogation in the vicinity of very low ($n \lesssim 15$) Rydberg states.  Conversely, microwave fields at the V/m level, and arbitrary frequency, could be measured using high-$n$ Rydberg states ($n\sim 100$).  Quantitative limits of these ranges will depend in part on ionization processes, Rydberg interactions, and many-body effects present in the system, for which further investigation is needed.  To achieve higher field measurement precision smaller cell sizes can be implemented for higher spectral resolution.  Lock-in detection and modulation spectroscopy for higher signal-to-noise could also be used~\cite{Yuechun.2017,Xiao.2000,Hodgekinson.2013}.

\section{Acknowledgements}
This material is based upon work supported by the Defense Advanced Research Projects Agency (DARPA) and the Army Contracting Command - Aberdeen Proving Ground (ACC-APG) under Contract number W911NF-15-P-0032.  The authors thank C.L. Holloway at NIST for technical support.

%

\end{document}